\documentclass[twocolumn,showpacs,nofootinbib]{revtex4}
\usepackage{amssymb}
\usepackage{graphicx}
\usepackage{dcolumn}

\def\be{\begin{equation}}
\def\ee{\end{equation}}
\def\bea{\begin{eqnarray}}
\def\eea{\end{eqnarray}}
\input epsf
\begin{document}

\title{{\LARGE The Curvaton Web}}
\author{Andrei Linde$^{1}$ and Viatcheslav Mukhanov$^2$}
\affiliation{$^1$Physics Department, Stanford University, Stanford CA 94305-4060, USA\\
$^{3}$ASC, Physics \ Department, LMU, Theresienstr. 37, Munich, Germany}

\begin{abstract}
We discuss nontrivial features of the large scale structure of the universe
in the simplest curvaton model proposed in our paper astro-ph/9610219. The
amplitude of metric perturbations in this model takes different values in
different parts of the universe. The spatial distribution of the amplitude
looks like a web consisting of exponentially large cells. Depending on the
relation between the cell size $\lambda_{0}$ and the scale of the horizon $l_{H}$, one may either live in a part of the universe dominated by gaussian perturbations
(inside a cell with $\lambda_{0} \gg l_{H}$), or in the universe dominated by nongaussian perturbations (for $\lambda_{0} \ll l_{H}$). We show that the curvaton
contribution to the total amplitude of adiabatic density perturbations can
be strongly suppressed if the energy density of the universe prior to the
curvaton decay was dominated not by the classical curvaton field but by the
curvaton particles produced during reheating. We describe the curvaton-inflaton transmutation effect: The same field in different parts of the universe may play either the role of the curvaton or the role of the inflaton.   Finally, we discuss an interplay between the curvaton web and anthropic considerations in the string theory landscape.
\end{abstract}

\maketitle
\tableofcontents

%\vskip 1cm

\section{Introduction}

It is well known that quantum fluctuations of the inflaton field in the
simplest inflationary models produce gaussian adiabatic perturbations of
metric with nearly scale independent spectrum \cite{Pert,muk1,mukhbook}.
This prediction is in a very good agreement with the present observational
data \cite{WMAP,Boom}. However, since the precision of the observational
data improves every year, one may try to discriminate between various
inflationary models by considering subtle deviations from the standard
paradigm. One way to distinguish between various models is by studying a
small degree of nongaussianity of perturbations, and a possible small
contribution of the isocurvature perturbations.

It is impossible to produce isocurvature perturbations in the models
describing a single scalar field, the inflaton, and the degree of
nongaussianity of the inflaton perturbations typically is extremely small.
However, it is relatively easy to produce isocurvature and/or nongaussian
adiabatic perturbations in the models containing several light scalar
fields. The first model of that type was proposed in \cite{Ax}. It was based
on investigation of axions, which were massless during inflation. At that
stage, the axion perturbations practically did not contribute to the
perturbations of curvature. At the late stages of the evolution of the
universe, the axion field became massive, and its initial isocurvature
perturbations were converted into a mixture of adiabatic and isocurvature
perturbations, which could also have a significant nongaussian component.
Another model of that type describing a mixture of adiabatic and
isocurvature perturbations was considered in \cite{Mollerach:1989hu}.

However, current observational data put very strong constraints on the
possible admixture of isocurvature perturbations \cite%
{WMAP,Gordon:2002gv,Boom}. Therefore recently a special attention was payed
to a class of models where a second field, called
``curvaton,'' at a certain stage begins to dominate and then
decays leading essentially to a secondary stage of reheating. In this model
the initial isocurvature fluctuations can be completely converted to the
adiabatic ones. The spectrum of perturbations in such models can be nonflat,
it can have various nonstandard features, and it can be nongaussan. This
possibility was first discussed in our paper \cite{LM}, and then it was
significantly developed in the series of papers by Lyth, Wands \cite{LW},
Moroi and Takahashi \cite{Moroi:2001ct} and other authors. A closely related
but different mechanism was also proposed in \cite{mod1}.

At the first glance, our original model \cite{LM} may seem extremely simple.
One of its versions describes a light field $\sigma$, which fluctuates
during inflation, and later begins to dominate the energy density of the
universe and decays. However, the simplicity of this model, as well as of
many subsequent versions of the curvaton scenario, is somewhat deceptive.
Such models must satisfy many specific conditions on the mass of the
curvaton, on the way it decays, on the way the CMB and the baryon asymmetry
of the universe are produced, etc., see e.g. \cite%
{LW,Gordon:2002gv,Bassett:2005xm}. Also, the predictions of the curvaton
scenario may depend on the choice of initial conditions and on our own
position in the universe. This last fact, which was mentioned in \cite{LM},
will be the main subject of our paper.

Most of the papers on the curvaton scenario discuss the behavior of the curvaton perturbations as a function of the classical nearly homogeneous curvaton field. Meanwhile we would like to concentrate on the global structure of the universe in this scenario, in which the curvaton field may take different values in different pars of the universe due to stochastic processes during inflation.  We will show that the spatial distribution of the amplitude of density perturbations in the curvaton scenario \cite{LM} looks like a web consisting of exponentially large cells. We will call this cell structure `the curvaton web.' Depending on the values of the parameters of the theory, the amplitude of density perturbations reaches its maximal values either in the internal parts of the cells, or near their boundaries.

We will also describe a potentially important effect, which, to the best of
our knowledge, was ignored in the existing literature on the curvaton
perturbations. It is often assumed that the curvaton energy density is
entirely due to the initial classical curvaton field and its inflationary perturbations, which also behave essentially as classical fields.
However, part of the inflaton energy can be converted to the energy of curvaton
particles during reheating after inflation. We will show that under certain
conditions the energy density of these particles can be much greater than
the energy density of the primordial curvaton field. As a result, this
effect may reduce the curvaton contribution to the total amplitude of
density perturbations. It may also allow to increase the degree of nongaussianity of these perturbations.

Finally, we will discuss the curvaton-inflaton transmutations and a possible interplay between the curvaton web and anthropic arguments related to the concept of string theory landscape.

\section{Generation of perturbations}

\label{pert}

Consider a theory of a scalar field $\sigma $ with a mass $m$, which gives a
subdominant contribution to the energy density during inflation. If $m$ is
much smaller than the Hubble constant $H$ during inflation, the
long-wavelength perturbations of this field with an average amplitude $%
\delta \sigma ={\frac{H}{2\pi }}$ are generated during each time interval $%
H^{{-1}}$. Ignoring the spatial variations of the classical field $\sigma $,
which are locally unimportant but can be significant on a sufficiently large
scale (see below), one finds that these perturbations give rise to \textit{%
locally gaussian} perturbations of density $\delta \rho _{\sigma
}=m^{2}\sigma \delta \sigma \sim {\frac{m^{2}\sigma H}{2\pi }}$. Their
relative amplitude compared to the (local) value of the energy density of
the field $\sigma $ is given by 
\begin{equation}
{\frac{\delta \rho }{\rho _{\sigma }}}={\frac{2\delta \sigma }{\sigma }}\sim 
{\frac{H}{\pi \sigma }}\ .  \label{locgauss}
\end{equation}%
This simple equation is valid only for $\delta \sigma > \sigma$; ${\frac{\delta \rho }{\rho _{\sigma }}}$ become $O(1)$ for $\sigma \lesssim H/2\pi$.

The curvaton scenario is based on the assumption that the perturbations ${%
\frac{\delta \rho }{\rho _{\sigma }}}$ can be significant. However, since
this field is subdominant during inflation and shortly after it, the
contribution of the curvaton fluctuations to the adiabatic perturbations of
metric related to ${\frac{\delta \rho }{\rho _{\mathrm{total}}}}$ is much
smaller than ${\frac{\delta \rho }{\rho _{\sigma }}}$: 
\begin{equation}
{\frac{\delta \rho }{\rho _{\mathrm{total}}}}={\frac{m^{2}\sigma \delta
\sigma }{\rho _{\mathrm{total}}}}\sim {\frac{m^{2}\sigma }{6\pi H}}\ ,
\label{locgausstot}
\end{equation}%
where we use the system of units $M_{p}^{2}\equiv \left( 8\pi G\right) ^{-1}
= 1.$

After inflation, the inflaton field decays and produces matter. If the
products of its decay are ultrarelativistic particles, then their energy
density subsequently decreases as $a^{-4}$, where $a$ is the scale factor of
the universe. Meanwhile, if the field $\sigma $ has a quadratic potential,
then for a long time its energy density decreases more slowly. It does not
decrease at all until the Hubble constant becomes smaller than $m$, and then
it decreases as $a^{-3}$. As a result, the field $\sigma $ may give the
dominant contribution to the energy density of the universe, and its
perturbations may give rise to significant perturbations of metric. If cold
dark matter and baryon asymmetry are produced during a secondary stage of
reheating due to the curvaton decay, then the adiabatic perturbations of
density are generated, without any admixture of isocurvature perturbations 
\cite{LM,LW}. These perturbations are of the order ${\frac{\delta \rho }{%
\rho _{\sigma }}}$. In principle, they can be much greater than the adiabatic
perturbations existing at the end of inflation.

Two comments are in order here. First of all, particles $\sigma $ can be
produced during the inflaton decay at the first stage of reheating. After
the moment when they become nonrelativistic, their energy density decreases
in the same way as the energy density of an oscillating homogeneous field $%
\sigma $, and of the long-wavelength fluctuations of the curvaton field.
This is an important and rather generic effect, which, to the best of our
knowledge, was not discussed in the literature. It can increase the total
energy density of the curvaton field at the time of its decay, and decrease
the resulting amplitude of the adiabatic perturbations produced by the
decaying curvaton field, see Section \ref{reh}.

Another comment concerns the origin of the classical scalar field $\sigma $
in the denominator of Eq. (\ref{locgauss}). This field can be nearly
homogeneous in the whole universe, with the value determined by initial
conditions at the beginning of inflation. However, if the mass of the field
is much smaller than the Hubble constant during inflation, more interesting
situation can be realized. Namely, the fluctuations of the curvaton field
produced during (eternal) inflation may bring the light field $\sigma $ to
different places in different parts of the universe. As a result, the
quantity ${\frac{\delta \sigma }{\sigma }}$ and the resulting amplitude of
adiabatic perturbations of metric take different values in different
exponentially large parts of the universe. This effect, as well as the
similar effect considered in \cite{Linde:1989tz} in the context of the
inflationary Brans-Dicke cosmology, provides a possible justification to the
anthropic considerations involving variable amplitude of perturbations of
metric \cite{Tegmark:1997in}.

In this paper we will describe a simplest version of the curvaton scenario
proposed in \cite{LM}, which is sufficient to explain our main idea. In the beginning, we will assume, for simplicity, that inflation occurs for a very long time, and
during this time the Hubble constant $H$ and the effective mass of the
curvaton field practically did not change. This is a reasonable assumption
in new inflation and in some versions of hybrid inflation. In
the subsequent sections we will consider possible generalizations of these
assumptions.

Under the specified conditions, the scalar field $\sigma$ during the long
inflationary evolution will gradually roll down towards $\sigma = 0$, and
then it will jump in all possible directions due to inflationary
perturbations. In this respect our simplest model \cite{LM} differs from the
scenario where the universe is filled by a homogeneous curvaton field.  As we will see in Section \ref{chaotic}, a generalized version of our scenario in fact describes a collection of exponentially large domains with all possible initial conditions for the field $\sigma$. However, if
one calculates the average value of the field $\sigma$ over the whole
universe in our scenario, one finds $\langle\sigma\rangle = 0$. Meanwhile,
for $\langle \sigma^2 \rangle$ in the simplest model with $H = const$ one finds the well-known Bunch-Davies
expression \cite{fluct,mukhbook,book}: 
\begin{equation}  \label{1ae}
\langle \sigma^2 \rangle = {\frac{H^2}{4\pi^2}} \int\limits_0^H {\frac{dk}{k}%
}\left({\frac{k}{H}}\right)^{\frac{2m^2}{3 H^2}} = {\frac{3H^4}{8 \pi^2 m^{2}%
}} \ .
\end{equation}
The estimate of the energy density of the fluctuations of the field $\sigma$
can be obtained as follows: 
\begin{equation}  \label{1ab}
\rho_\sigma \simeq {\frac{ m^2 \langle \sigma^2 \rangle}{2}} = {\frac{3H^4}{%
16 \pi^2}} \ .
\end{equation}

To get an intuitive understanding of this result, one should take into
account that the fluctuations of the field $\sigma$ with mass $m^{2} \ll
H^{2}$ produced during each time interval $\Delta t = H^{{-1}}$ have an
average amplitude $\delta\sigma = {\frac{H}{2\pi}}$ \cite%
{fluct,mukhbook,book}. In the beginning, an average square of the amplitude
of these fluctuations grows as $\langle\sigma^{2}\rangle = {\frac{H^{3}}{%
4\pi^{2}}}t$, just like in the theory of a massless scalar field. However,
the amplitude of long-wavelength fluctuations of a massive scalar field
decreases as $\sigma(t) = \sigma_{0} \exp \left(-{\frac{m^{2}t}{3H}}\right)$. Therefore the amplitude of the fluctuations will start decreasing
exponentially after the time $\Delta t \sim H/m^{2}$ since the moment they
were produced. As a result, the main contribution to $\langle\sigma^{2}%
\rangle$ will be given by perturbations produced during the latest time
interval $\Delta t \sim {\frac{3H}{m^{2}}}$, because all other perturbations
will become exponentially small. This gives the estimate $%
\langle\sigma^{2}\rangle \sim {\frac{H^{3}}{4\pi^{2}}} \times {\frac{3H}{%
m^{2}}} = {\frac{3H^4}{4 \pi^2 m^{2}}}$, which almost coincides with the
Bunch-Davies result.

Note that inflationary fluctuations with $k\ll H$ correspond to quantum
fluctuations with large occupation numbers $n_{k}={\frac{H^{2}}{k^{2}}}\gg 1$
\cite{book}. Therefore these fluctuations can be interpreted as waves of a
classical scalar field. The quantity 
\begin{equation}
\sqrt{\langle \sigma ^{2}\rangle }=\sqrt{\frac{3}{2}}{\frac{H^{2}}{2\pi m}}
\label{1ac}
\end{equation}%
represents an average deviation of the inhomogeneous classical scalar field
from zero on an exponentially large scale $l\ll \lambda _{0}$, to be discussed in the next section.

\section{The curvaton web}

\label{web}

If we decide, as it is commonly done, to concentrate our attention only on $%
\langle \sigma \rangle =0$ and $\langle \sigma ^{2}\rangle ={\frac{3H^{4}}{%
8\pi ^{2}m^{2}}}$, or on any other averages involving an arbitrary number of
operators $\sigma $ at different points, we will be unable to get a full
picture of the distribution of perturbations within relevant for
observations regions. In order to go beyond the simple perturbative
approach, one should first remember that as a result of accumulation of the
long-wavelength quantum fluctuations of the field $\sigma $, the universe
becomes divided into many exponentially large regions filled with the 
\textit{positive} field $\sigma $ with the typical value $\sigma \sim +{%
\frac{H^{2}}{2\pi m}}$, and many exponentially large regions containing 
\textit{negative} field $\sigma $ with the typical value $\sigma \sim -{%
\frac{H^{2}}{2\pi m}}$.

\begin{figure}[h!]
\hskip -1cm \epsfysize=7 cm \epsfbox{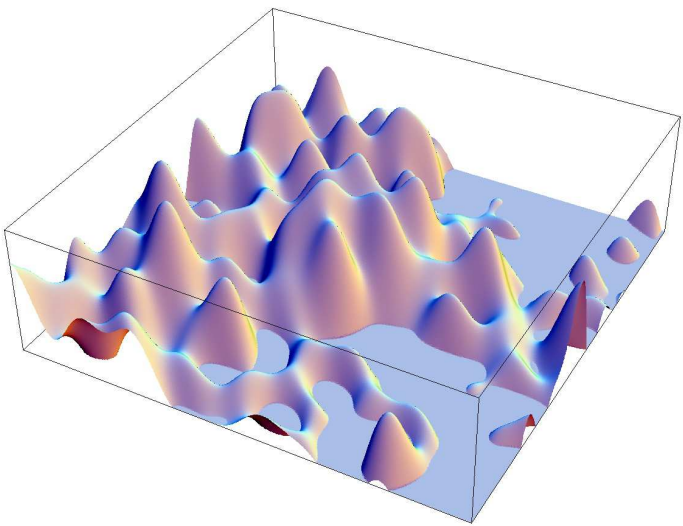}
\par
\ 
\caption{Results of a computer simulation of the Bunch-Davies distribution
of the curvaton field $\protect\sigma$ for $H/m = 3$. For greater values of $%
H/m$ the typical height of the hills grow as $H^{2}/m$, and a typical size
of the `islands' grows as $H^{-1} \exp\Bigl(O\Bigl({\frac{H^2}{m^2}}\Bigr)\Bigr)$. In
order to make the level $\protect\sigma=0$ clearly visible, we show only the
part of the `curvaton landscape' with $\protect\sigma > 0$, i.e. above the
`sea level' $\protect\sigma = 0$. As we will show below, the amplitude of
the curvaton density perturbations behaves in a rather peculiar way near the
`shoreline' $\protect\sigma = 0$, see Figs. \protect\ref{fig:Figa} and 
\protect\ref{fig:Figb}. }
\label{fig:Fig01}
\end{figure}

In order to understand the properties of the ``curvaton landscape'' one should take into account that for the observers interested in the perturbations on the scale $l \sim k^{{-1}}$, all perturbations with the wavelength $\lambda \gg k^{-1}$ will look like
a homogeneous field. For these perturbations, which look locally homogeneous
on a scale $k^{-1}$, we have 
\begin{equation}  \label{1ae2}
\langle \sigma^2 \rangle_{k} = {\frac{3H^4}{8\pi^2m^{2}}} \left({\frac{k}{H}}%
\right)^{\frac{2m^2}{3 H^2}}\ .
\end{equation}
One can easily check  that 50\% of the total Banch-Davies value $ {\frac{3H^4}{8\pi^2m^{2}}}$ is given by perturbations which look relatively homogeneous on scale smaller than $\lambda_{0.5} \sim H^{-1} \exp\Bigl({\frac{H^2}{m^2}}\Bigr)$
and 90\% of the Banch-Davies value is given by perturbations which look homogeneous on scale smaller than $\lambda_{0.9} \sim H^{-1} \exp\Bigl({\frac{H^2}{6m^2}}\Bigr)$. Note that the scales $H^{-1} \exp\Bigl({\frac{H^2}{m^2}}\Bigr)$ and  $H^{-1} \exp\Bigl({\frac{H^2}{6m^2}}\Bigr)$ are exponentially different from each other.   Therefore it would be incorrect to think that the universe after inflation looks like a chessboard with cells of positive and negative $\sigma$ of approximately equal size $l \sim \lambda_{0.5}$ (or $l \sim \lambda_{0.9}$).  Still the logarithms of the scales $\lambda_{0.5}$ and  $\lambda_{0.9}$ are similar and proportional to $H^{2}/m^{2}$. In this rather vague sense we will be talking about the distribution of the curvaton field being locally homogeneous on scale
\begin{equation}
\lambda_0 \sim H^{-1} \exp\Bigl(O\Bigl({\frac{H^2}{m^2}}\Bigr)\Bigr)\ .
\end{equation}
To get a better understanding of the spatial distribution of the curvaton field we performed computer simulations of the process of formation and accumulation of the curvaton perturbations, using the methods developed in Ref. \cite{LLM}, see Fig.~\ref{fig:Fig01}.

Exponentially large  domains with positive and negative $\sigma$ are separated by the boundaries with $\sigma = 0$, which correspond to the `shoreline' in Fig.~\ref{fig:Fig01}. The 3D distribution of the domain boundaries with $\sigma = 0$, which we will call `the curvaton web,' resembles the network of domain walls in the theories with spontaneous breaking of a discrete symmetry. The main difference is that usually the domain walls have large energy density; they exist only because of the
topological reasons. In our case the situation is opposite: The potential
energy density of the field $\sigma$ is of the order $H^{4}$ inside the
domains, and it vanishes at the boundaries with $\sigma =0$. On the other hand, the amplitude of the curvaton density perturbations behaves in a rather peculiar way near the walls with $\sigma = 0$, see Figs. \protect\ref{fig:Figa} and 
\protect\ref{fig:Figb}.

In order to evaluate the perturbations of metric, let us first consider
perturbations on the scale much larger than the size of a typical
domain, $l\sim k^{{-1}}\gg \lambda _{0}$ \cite{LM}. In this case the fluctuating field $\sigma $
wanders many times in the region $-H^{2}/m{\ \lower-1.2pt\vbox{\hbox{%
\rlap{$<$}\lower5pt\vbox{\hbox{$\sim$}}}}\ }\sigma {\ \lower-1.2pt%
\vbox{\hbox{\rlap{$<$}\lower5pt\vbox{\hbox{$\sim$}}}}\ }H^{2}/m$ on the
scale $l$, so its value averaged over the domain of a size $\sim k^{{-1}}$
vanishes. As a result, addition of a perturbation $\delta \sigma (k)$ with $%
k\ll k_{0}$ does not lead to the usual density perturbations $m^{2}\sigma
\delta \sigma $ \cite{Lyth}. Perturbations of density will be nongaussian. To get a rough estimate of their amplitude, one may assume that they are quadratic in $\delta \sigma $, which would lead to an estimate  for ${\delta \rho\over\langle\rho_{\sigma}\rangle}$ at the stage of inflation,
\begin{equation}
{\delta \rho\over\langle\rho_{\sigma}\rangle}\sim {\frac{m^{2}}{H^{2}}}\left( {\frac{k}{H}}\right) ^{%
\frac{2m^{2}}{3H^{2}}},  \label{6i}
\end{equation}%
see Eq. (5) in \cite{LM}. (Note that here we are writing $\langle\rho_{\sigma}\rangle = O(H^{4})$ instead of the local value $\rho_{\sigma}(x)$.) However, it was pointed out in our paper \cite{LM} that for a more accurate evaluation of the amplitude of the perturbations one should find
the correlation function 
\begin{eqnarray}
\xi (r, t)=\frac 1{\left\langle \rho _\sigma \right\rangle ^2}\left(
\left\langle \rho _\sigma (\stackrel{\rightarrow }{x}, t)\rho
_\sigma (\stackrel{\rightarrow }{x}+\stackrel{\rightarrow }{r}%
, t)\right\rangle -\left\langle \rho _\sigma \right\rangle ^2\right) .
\label{corr}
\end{eqnarray}
After that one can find ${\delta \rho\over\langle\rho_{\sigma}\rangle}$ from the relation
\[
\xi \left( r, t\right) =\int \frac{dk}k\frac{\sin kr}{kr}\left( {\delta \rho\over\langle\rho_{\sigma}\rangle} \right) ^2.
\]
This yields \cite{LM}
\begin{equation}
\left( {\delta \rho\over\langle\rho_{\sigma}\rangle} \right) ^2\sim \frac{m^4k^3}{\left\langle
\rho _\sigma \right\rangle ^2}\int \left| \sigma _{\stackrel{\rightarrow
}{k^{\prime }}}\right| ^2\left| \sigma _{\stackrel{\rightarrow
}{k}-\stackrel{\rightarrow }{k^{\prime }}}\right| ^2d^3 k \ ,
\label{amp}
\end{equation}
where $\sigma_{k}^{2}k^{3} = {H^{2}\over   2} \left(k^{2}\over H^{2}\right)^{2m^{2}/3H^{2}} $ for $k \ll H$. Because we are
only interested in large scale fluctuations produced at
inflation, we can make a UV cut off at $k\sim  H$.  The corresponding calculation have been made in \cite{LM} taking into account the complicated inflationary and post-inflationary evolution in the chaotic inflation scenario for all wavelengths. However, in the context of the present discussion we are assuming that the Hubble constant during inflation remains constant, and we will be interested only in the long wavelengths which do not loose their gradient energy due to the subsequent redshift. In this case the calculation is much simpler than the calculation performed in our paper \cite{LM}, but the result is qualitatively the same as Eq. (17) of Ref. \cite{LM}:
\begin{equation}
 {\delta \rho\over\langle\rho_{\sigma}\rangle} \sim {m\over H} \left( \frac {k}{H}\right)  ^{\frac{2m^{2}}{3H^{2}}}  \ .   \label{spe1}
\end{equation}
 Note that $\rho_{\sigma}$ and the perturbations $\sigma_{k}^{2}$ scale in the same way during the subsequent expansion of the universe (they decrease during the field oscillations as $a^{{-3}}$), this equation remains valid until the perturbations enter the horizon. 
 
 The reason why this result is greater than the naive estimate (\ref{6i}) by a factor of $H/m$ is pretty simple. The estimate (\ref{6i}) was obtained by integration of the square of perturbations $\delta \sigma \sim H/2\pi$. However, the positive (negative) perturbations $\delta\sigma$ are correlated with the exponentially large domains where the field $\sigma$ is positive (negative) and takes average value $\sim H^{2}/m$, which is $H/m$ times greater than $\delta\sigma$.

Eq. (\ref{spe1}) implies that these
perturbations have blue spectrum. Note, however, that one can easily obtain
red spectrum as well, or a spectrum of a more complicated shape, if one
takes into account a slow decrease of $H$ and a possible change of $m$
during inflation.

We should note also that Eq. (\ref{spe1}) describes the ratio  ${\delta \rho\over\langle\rho_{\sigma}\rangle}$, where $\langle\rho_{\sigma}\rangle$ is the {\it average} energy density of the field $\sigma$. On scale $l\gg \lambda _{0}$ the local energy density $\rho_{\sigma}$ changes substantially, so in some places, e.g. near the walls of the curvaton web, where $\sigma = 0$, the quantities  ${\delta \rho\over\langle\rho_{\sigma}\rangle}$ and  ${\delta \rho\over \rho_{\sigma}}$ may be quite different from each other.

Now let us consider an opposite limit: $l\ll \lambda _{0}\sim H^{-1} \exp\Bigl(O\Bigl({\frac{H^2}{m^2}}\Bigr)\Bigr)$. On this scale the value of the field $%
\sigma $ remains practically constant, and the relative perturbations of
density of the field $\sigma $ are (locally) gaussian, with an amplitude
given by Eq. (\ref{locgauss}): 
\be\label{yyyy}
{\frac{\delta \rho }{\rho _{\sigma }}}={\frac{2\delta \sigma }{\sigma }}\sim 
{\frac{H}{\pi \sigma }}\left( \frac {k}{H}\right)  ^{\frac{m^{2}}{3H^{2}}}. 
\ee
The last term appears due to the red tilt of the fluctuations, but this term is indistinguishable from 1 for $k^{-1} \sim l \ll \lambda_{0}$. 

This equation shows that ${\frac{\delta \rho }{\rho _{\sigma }}}$ can be
large at small $\sigma $. Strictly speaking, this equation is not correct in
the regions near $\sigma =0$. Indeed, the field $\sigma $ deviates from $%
\sigma =0$ by $\delta \sigma \pm {\frac{H}{2\pi }}$ on the scale $H^{-1}$,
which means that the relative amplitude of density perturbations becomes
large near $\sigma =0$: \ ${\frac{\delta \rho }{\rho _{\sigma }}}=O(1)$.

A typical amplitude of perturbations can be obtained by substitution of the
average amplitude of $\sigma $ from Eq. (\ref{1ac}): $\sigma \sim \sqrt{%
\frac{3}{2}}{\frac{H^{2}}{2\pi m}}$. This gives 
\begin{equation}
{\frac{\delta \rho }{\rho _{\mathrm{\sigma }}}}\sim {\frac{m}{H}}\ .
\label{locgausstot2ab}
\end{equation}
Thus, up to a factor $\left( \frac {k}{H}\right)  ^{\frac{m^{2}}{3H^{2}}}$, which is approximately equal to 1 for $l\ll \lambda _{0}$, the typical amplitude of the fluctuations on scale $l\ll \lambda _{0}$ (\ref{locgausstot2ab}), even though the nature and interpretation of these perturbations is quite different. Meanwhile on scale $l\gg \lambda _{0}$ the amplitude of perturbations  (\ref{spe1}) becomes suppressed by the factor $ \left( \frac {k}{H}\right)  ^{\frac{2m^{2}}{3H^{2}}}$.

On the other hand, if one is interested in the local amplitude of density perturbations for $l \ll \lambda_{0}$ (which is the case if $\lambda_{0}$ is much greater than the scale of the observable part of the universe) one should use Eq.  (\ref{yyyy}).

\section{Curvaton production during reheating}\label{reh}

The effect which we are going to consider in this section can be very essential, but
it is model-dependent. In order to make our consideration as simple as
possible, we will restrict ourselves to the order of magnitude estimates and
assume that at the last stages of inflation the Hubble constant $H,$ the
inflaton mass $M$ and the curvaton mass $m$ are roughly of the same order of
magnitude. In reality, we expect that $M\gg H\gg m$, but this does not alter
our main conclusion. In fact, it is very easy to make our consideration more
general and precise, following \cite{DL,KLS}, but we do not want additional model-dependent complications to distract us from our main qualitative conclusion.

Our main assumption is that the inflaton field is in the hidden sector, and
decays to all other fields only due to gravitational effects, with the decay
rate $\Gamma \sim M^{3}$ (in Planck units $M_{p} = 1$). We will also assume
that the curvaton field interacts only very weakly with itself and other
particles, so that nonrelativistic curvaton particles produced as a result
of inflaton decay do not thermalize, and decay at the same time as the
classical curvaton field and its long-wavelength fluctuations generated on inflation.

At the end of inflation, the energy density of the inflaton field $\phi $ is
of order $H^{2}$, while the energy of the fluctuations of the curvaton field $\sigma $ is $H^{4}$. After inflation, the energy density of the inflaton is decreasing, while
the curvaton energy density remains constant until the time $t\sim m^{{-1}}$, when the long wavelength curvaton field  begins to oscillate. At that time, the inflaton energy density is
of order $m^{2}$. Starting from this moment, the ratio of the inflaton
and the curvaton energy densities remains constant, until the inflaton
decays. However, when $m$, $M$, and $H$ are not different, this is a minor
effect. In this case, both inflaton and curvaton field begin to oscillate
nearly immediately after the end of inflation when their energy densities
are $H^{2}$ and $H^{4}$ respectively, and their ratio, $\sim H^{-2},$ is
preserved until the inflaton decays.

Ignore for a moment that the inflaton decays to many different types of
particles and suppose that its energy goes entirely to the curvaton
particles. Then the energy density of the curvaton particles will be greater
than the energy density of the long-wavelength perturbations of the curvaton
field by the factor $H^{-2}$. Note that the constraints on the amplitude of
gravitational waves during inflation implies that $H<10^{-5}$, so that $%
H^{-2}$ is extremely large: $H^{-2}>10^{10}$, in Planck units. That is why
we were ignoring numerical factors, the difference between $H$, $M$ and $m$,
etc. The main effect will be present even if we relax many of our
assumptions.

Now let us use our estimates for calculating the curvaton contribution to
the amplitude of adiabatic perturbations. This amplitude is given by ${\frac{%
\delta \rho }{\rho _{\sigma }}}$, where $\delta \rho =m^{2}\sigma \delta
\sigma $. As for $\rho _{\sigma }$, in all papers on this subject this
quantity is supposed to be either the energy density of the initial
classical field $\sigma $ or its generated long wavelength perturbations. In
our scenario the typical value of this last contribution is of order $%
m^{2}\left\langle \sigma ^{2}\right\rangle \sim H^{4}$. Assuming that it
dominates and, hence, $\sigma \sim H^{2}/m,$ we obtain ${\frac{\delta \rho }{%
\rho _{\sigma }}}\sim {\frac{m}{H}}$, as in Eq. (\ref{locgausstot2ab}). In
order to have ${\frac{\delta \rho }{\rho _{\sigma }}}\sim 10^{-5}$ one needs
either $m\sim 10^{{-5}}H$ \ or, alternatively, we may assume, as many people
do, that the initial value of the classical curvaton field was much larger
than $H^{2}/m,$ thus, reducing the amplitude of perturbations.

Now let us take into account the curvaton production during reheating. First
note that the total energy density of the produced curvaton particles can
exceed the energy density of the long-wavelength fluctuations by the factor
of $H^{-2}$, and therefore the resulting amplitude of perturbations is
reduced by the factor of $H^{2}.$ Thus even if $m$ is comparable with $H$,
we can obtain nevertheless 
\begin{equation}
{\frac{\delta \rho }{\rho _{\sigma }}}\sim H^{2}\ .  \label{locgausstot11}
\end{equation}%
For $H<10^{-5}$, this quantity is smaller than $10^{{-10}}$. It is well
below than the level of the standard adiabatic perturbations produced by the
inflaton field and even much smaller than the amplitude of the tensor
perturbations $\sim H$. In other words, the $\sigma $-particles production
during reheating can dramatically reduce the curvaton contribution to the
final amplitude of adiabatic density perturbations making it negligibly
small.

As we have noted above, one can also decrease the contribution of the
curvaton perturbations into the final amplitude of the resulting adiabatic
fluctuations by considering the parts of the universe with a very large
classical curvaton field $\sigma $. However, if $\sigma > 1$, then the field $\sigma$ will drive chaotic inflation and will play the role of the inflaton instead of the curvaton; see more about it in Section \ref{chaotic}. On the other hand, one can easily check that if $\sigma <1$ and one ignores production of curvaton particles during  reheating, then the amplitude of the curvaton contribution ${\frac{\delta \rho }{\rho _{\sigma }}}={\frac{2\delta \sigma }{\sigma }}\sim 
{\frac{H}{\pi \sigma }}$ is always greater than the amplitude of tensor perturbations $\sim H$ produced during inflation.

The possibility of a strong suppression of the curvaton contribution due to reheating, Eq. (\ref{locgausstot11}),  does not mean that the final amplitude of the scalar metric
perturbations can be made smaller than the amplitude of the tensor
perturbations. If the contribution of the isocurvature curvaton
mode is too small, then the resulting amplitude of metric perturbations is determined by the usual adiabatic perturbations produced by the inflaton field \cite{Linde:2005he}.

As we already mentioned, the results obtained above are strongly model dependent. One can easily avoid suppression of the curvaton perturbations by changing the assumptions about the mechanism of reheating. The
main goal of this consideration was to show that in the calculations of the
curvaton perturbations, in addition to the energy of a classical curvaton
field and its long-wavelength inflationary perturbations one should also
consider the energy of the curvaton particles produced during reheating.
This contribution may be either negligible, or extremely important, as in
the case described above. Since the importance of the effect is
model-dependent, we will describe it phenomenologically by modifying the
equation for the curvaton contribution to the amplitude of (locally
Gaussian) density perturbations as 
\begin{equation}
{\frac{\delta \rho }{\rho _{\sigma }}}\sim {\frac{2m^{2}\sigma \delta \sigma 
}{m^{2}(\sigma ^{2}+C^{2})}}\sim {\frac{\sigma H}{\pi (\sigma ^{2}+C^{2})}}\
.  \label{locgausstot11a}
\end{equation}%
Here $\sigma $ is the locally measured amplitude of the classical curvaton
field and $C$ is a phenomenological parameter proportional to the energy
density of the curvaton particles produced during reheating.

For $C=0$ and $\sigma \sim {\frac{H^{2}}{2\pi m}}$ this equation reproduces
our previous result ${\frac{\delta \rho }{\rho _{\mathrm{\sigma }}}}\sim {%
\frac{m}{H}}$, but as we have seen above, under certain conditions the
constant $C$ may be quite large, which leads to a decrease of the amplitude
of the curvaton contribution to density perturbations, and to a different
dependence of this amplitude on the local value of the field $\sigma $.
For $C\not=0$, the amplitude of the density perturbations does not blow up
at $\sigma =0$, as before. It reaches its maximum at $\sigma =C$: 
\begin{equation}
{\frac{\delta \rho }{\rho }}_{\mathrm{max}}\sim {\frac{H}{2\pi C}}\ .
\label{locgausstot11ab}
\end{equation}

One of the interesting consequences of this result is that even if one
ignores the effects of the curvaton web, one can still obtain strongly
nongaussian perturbations generated by the curvaton mechanism. Indeed,
usually the amplitude of the locally gaussian fluctuations was supposed to
be ${\frac{\delta \sigma }{\sigma }}\sim 10^{{-5}}$. The smallness of this
ratio implied that the level of nongaussianity was expected to be extremely
small (quadratic in ${\frac{\delta \sigma }{\sigma }}$). However, with an
account taken of the curvaton particles produced during reheating, one can
obtain a small amplitude of density perturbations even if ${\frac{\delta
\sigma }{\sigma }}\gg 10^{{-5}}$. This may substantially increase the level
of nongaussianity of adiabatic perturbations produced by the curvatons. In
other words, our ignorance of the specific mechanism of curvaton production
during reheating makes the degree of nongaussianity in the curvaton scenario
a free parameter.

Finally we should note that Eq. (\ref{locgausstot11a}) is valid in a more general case, if during the curvaton decay there were other particles, giving contribution $\rho_{p} = 2m^{2}C^{2}$ to the total energy density during the curvaton decay. We concentrated on the case where this contribution is dominated by the curvaton particles, since in this case  one does not get any undesirable isocurvature perturbations if CDM, baryon and lepton asymmetry are generated either by the curvaton decay or after it. One can also avoid the isocurvature perturbations even if the energy density during the curvaton decay is dominated by radiation, under the assumption that  CDM, baryon and lepton asymmetry are produced  after the curvaton decay, but not due to the curvaton decay \cite{Lyth:2002my}.

\section{Reheating and the curvaton web}

The observational consequences of our scenario depend on relations between
the typical size of the curvaton domain $\lambda _{0},$ the size of the
observable part of the universe $l_{H}$ and a typical galaxy scale $%
l_{g}\sim 10^{{-6}}l_{H}$ (recalculated to the end of inflation).

As we have seen in the previous section, effects of reheating may allow us
to have a small amplitude of locally gaussian density perturbations not only
for $H$ many orders of magnitude greater than $m$, but also for $H,$ which
is only a few times greater than $m$. This means that, depending on the
ratio between $H$ and $m$, a typical size of each domain with a positive or
negative $\sigma $, $\lambda _{0}\sim \exp\Bigl(O\Bigl({\frac{H^2}{m^2}}\Bigr)\Bigr)$, can take values either much smaller or much greater than the scale
of horizon $l_{H}$.

We will mention here two interesting regimes:

If $l_{H} \gg l_{g}\gg \lambda_{0}$, the perturbations are strongly
nongaussian, which contradicts observational data. On the other hand, if the
amplitude of these fluctuations is subdominant compared to the usual
adiabatic inflationary perturbations, then one can use this regime as
providing a small controllable amount of nongaussian isocurvature and
adiabatic perturbations \cite{LM}. The existence of the curvaton web helps
to understand the origin of this nongaussianity.

If $l_{H}\ll \lambda _{0}$, we live in a part of the universe dominated by
nearly gaussian adiabatic perturbations. For a proper choice of parameters,
such a possibility is consistent with all available observational data. A
nontrivial part of this scenario is that the amplitude of adiabatic
perturbations depends not only on the parameters of the model, but also on
the local value of the curvaton field at the end of inflation, which can
vary from $0$ to $H^{2}/m$ in different parts of the universe. This helps to
justify anthropic considerations comparing the properties of the universe
with different values of the amplitude of density perturbations. In this
respect, this scenario is similar to the scenario studied in \cite%
{Linde:1989tz}, where the amplitude of density perturbations could take
different values in different parts of the universe due to inflationary
perturbations of the Brans-Dicke field. Recently this possibility was
studied in \cite{Garriga:2005ee} in relation to the anthropic principle and
string landscape scenario. We will return to the discussion of this issue in
Section \ref{landscape}. The picture of the curvaton web allows not only to
discuss various probability distributions, but also to have a visual
understanding of the structure of the universe in this scenario.

The details of this picture depend on creation of $\sigma $-particles during reheating discussed in the previous section. In the case $C=0$ considered in Section \ref{web} (no curvaton particle production during reheating), the
maximum of density perturbations $\frac{\delta \rho }{\rho _{\sigma }}$ is
reached near the walls $\sigma =0$ separating the exponentially large
domains with positive and negative values of $\sigma $. However, for $%
C\not=0 $, the maximum would correspond to a finite value $|\sigma _{m}|=C$
(see Eq. (\ref{locgausstot11a})). If $C>H^{2}/m$, then the amplitude of
density perturbations will take its largest values deep inside these
domains, when $\sigma \sim H^{2}/m$, and it will be minimal near the walls
with $\sigma =0$. On the other hand, if $C<H^{2}/m$, then the amplitude of
density perturbations will be small inside the domains, it will grow on the
way towards the walls with $\sigma =0$, but then it will reach some maximal
value and vanish near the walls with $\sigma =0$. This means that the places
where the amplitude of density perturbations is maximal form two walls
surrounding from each side the wall with $\sigma =0$. Thus the effects of
reheating may affect in a rather nontrivial way the structure of the
curvaton web.

\begin{figure}[h!]
 \epsfysize=7.4 cm{\hskip -1cm\epsfbox{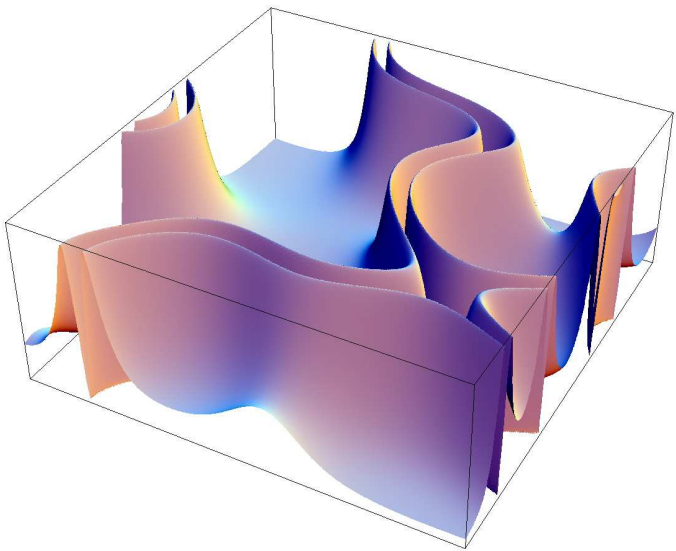}}
\par
\ 
\caption{The spatial distribution of the square of the amplitude of density
perturbations near the walls $\protect\sigma = 0$, for $C<H^{2}/m$. The amplitude
of density perturbations vanishes at the walls $\protect\sigma = 0$. These
walls are surrounded from both sides by the walls where the amplitude of
density perturbations reaches its maximal value.}
\label{fig:Figa}
\end{figure}

\begin{figure}[h!]
\centering\leavevmode\epsfysize=7.4 cm \epsfbox{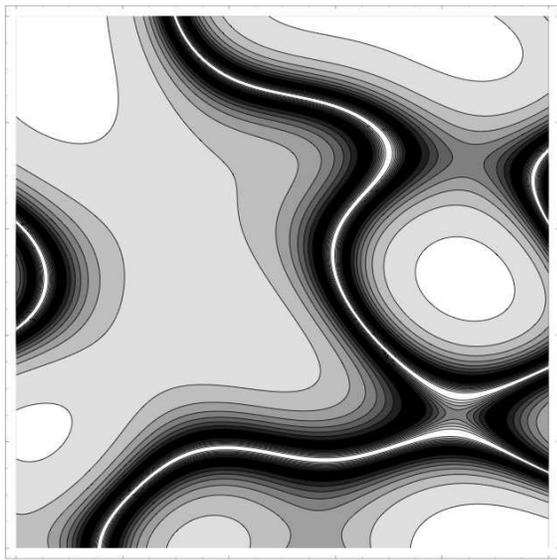}
\par
\ 
\caption{Contour plot for the square of the amplitude of the density
perturbations shown at the previous figure. White regions correspond to
vanishing amplitude of density perturbations, black regions correspond to
maximal absolute values of density perturbations. This figure confirms that
the two walls with large amplitude of density perturbations (see Fig.~\ref{fig:Figa}) surround the domain boundaries with $\protect\sigma %
= 0$ where the amplitude of density perturbations vanishes. (At the previous
figure, the regions with $\protect\sigma = 0$ were hidden by the walls.)}
\label{fig:Figb}
\end{figure}

Our results are illustrated by a set of two figures, which show the spatial
distribution of the square of the amplitude of the curvaton density
perturbations based on computer simulations of the curvaton web for $C<H^{2}/m$, see Fig.~\ref{fig:Figa} and Fig.~\ref{fig:Figb}. These two figures show why did we call the resulting structure `the
curvaton web.'

These figures clearly explain the main source of nongaussianity in the curvaton scenario of Ref. \cite{LM}. They also suggest a rather exotic possibility which may occur if we live not very far from the walls shown in Fig.~\ref{fig:Figa}. In this case the amplitude of density perturbations will be slightly different at the two opposite parts of the sky, depending on the distance from the nearby wall.

\section{Curvatons and eternal chaotic inflation}

\label{chaotic}

\subsection{Curvatons and chaotic inflation}

Until now we were assuming that the Hubble constant and the mass of the
curvaton field remain constant during inflation. This possibility can be
realized, for example, in new or hybrid inflation. In chaotic inflation the
Hubble constant changes significantly during inflation. Moreover, the
curvaton mass can depend on the curvature,  $m^{2}=m_{0}^{2}+\alpha
H^{2}$, see \cite{LM}. In this Section we consider how the results obtained in the previous sections are modified in this case. We will also assume, as in the previous sections, that initially the scalar field $\sigma $ was small; for a discussion of a
more general situation see \cite{LW,Hattori:2005ac} and also the end of section \ref{constant}.

We consider the model where inflation is driven by the field $\phi $ with an
effective potential $V(\phi )$ and assume that the curvaton mass is much
smaller than the Hubble constant during inflation, that is $m^{2}\ll H^{2}$.
In this case the long-wavelength modes of the curvaton field are in the
slow-roll regime and satisfy the equation%
\begin{equation}
3H\dot{\sigma}_{k}+m^{2}\sigma _{k}\simeq 0\ ,  \label{a1}
\end{equation}%
and, hence, for every mode with a given comoving wavenumber $k$ we have%
\begin{equation}
{\frac{d\sigma _{k}^{2}}{dt}=-}\frac{2m^{2}}{3H}\sigma _{k}^{2}\ .
\label{a2}
\end{equation}%
Integrating over $k$ and taking into account the contribution of the newly
generated from quantum fluctuation modes we obtain \cite{book}:%
\begin{equation}
{\frac{d\langle \sigma ^{2}\rangle }{dt}=-}\frac{2m^{2}}{3H}\langle \sigma
^{2}\rangle +{\frac{H^{3}}{4\pi ^{2}}}\ .  \label{a3}
\end{equation}%
The solution of this equation with the initial condition $\langle \sigma
^{2}\left( t=t_{i}\right) \rangle =0$ is 
\begin{equation}
\langle \sigma ^{2}\left( t\right) \rangle =\frac{1}{4\pi ^{2}}%
\int\limits_{t_{i}}^{t}H^{3}\left( \tilde{t}\right) \exp \left( -\frac{2}{3}%
\int\limits_{\tilde{t}}^{t}\frac{m^{2}}{H\left( \bar{t}\right) }d\bar{t}%
\right) d\tilde{t}\ .  \label{a4}
\end{equation}%
If $H$ and $m$ are both constant we obtain from here the familiar result (%
\ref{1ae}) as $t\rightarrow \infty .$ In the case of chaotic inflation it is
convenient to rewrite this equation using the value of the inflaton field $%
\phi $ as a time variable instead of $t.$ Taking into account that 
\begin{equation}
\frac{d\phi }{dt}\simeq -\frac{V^{\prime}}{3H},\text{ \ \ }H^{2}\simeq \frac{%
V}{3}\ ,  \label{a5}
\end{equation}%
where we again use the system of units $M_{p}=1$, equation (\ref{a4}) becomes%
\begin{equation}
\langle \sigma ^{2}\left( \phi \right) \rangle =\frac{1}{12\pi ^{2}}%
\int\limits_{\phi }^{\phi _{i}}\frac{V^{2}\left( \tilde{\phi}\right) }{V^{\prime}}\exp \left( -2\int\limits_{\phi }^{\tilde{\phi}}\frac{m^{2}}{%
V^{\prime}}d\bar{\phi}\right) d\tilde{\phi}\ ,  \label{a6}
\end{equation}%
where $\phi _{i}$ is the initial value of the inflaton field. Let us
consider separately two cases: $m^{2}=const$ and $m^{2}=\alpha H^{2},$ where 
$\alpha $ is some small constant, $\alpha <1$.

\subsection{ Curvaton with $m^{2}=const$}\label{constant}

\label{mconst} We begin with simple inflation due to the massive inflaton
field with potential $V=\frac{1}{2}M^{2}\phi ^{2}.$ In this case Eq. (\ref%
{a6}) implies that at the end of inflation, when $\phi \simeq 1,$ we have%
\begin{equation}
{\langle \sigma ^{2}\rangle }\simeq {\frac{M^{2}\phi _{i}^{4-\frac{2m^{2}}{%
M^{2}}}}{48\left( 4-\frac{2m^{2}}{M^{2}}\right) \pi ^{2}}.}  \label{a7}
\end{equation}%
In particular, for $m^{2}\ll M^{2}$ one obtains \cite{Felder:1999wt}: 
\begin{equation}
{\langle \sigma ^{2}\rangle }\simeq {\frac{M^{2}\phi _{i}^{4}}{192\pi ^{2}}}
\ .  \label{a8}
\end{equation}

These fluctuations lead to the same consequences as a classical scalar field 
$\sigma $ which is homogeneous on the scale $H^{-1}$ and which has a typical
amplitude 
\begin{equation}
\sigma _{0}=\sqrt{\langle \sigma ^{2}\rangle }={\frac{M\phi _{i}^{2}}{8\pi 
\sqrt{3}}} \ .  \label{8aaaa}
\end{equation}

As we see, the value of $\sigma _{0}$ depends on the initial value of the
field $\phi $. This result has the following interpretation. One may
consider an inflationary domain of initial size $H^{-1}(\phi _{i})$. This
domain after inflation becomes exponentially large. In the model with $%
V(\phi )={\frac{M^{2}}{2}}\phi ^{2}$ its size becomes \cite{book} 
\begin{equation}
\lambda _{0}\sim H^{-1}(\phi _{i})\exp \left({\phi _{i}^{2}\over 4}\right) .
\label{8aab}
\end{equation}%
One can easily check that the largest contribution to $\langle \sigma
^{2}\rangle $ is given by perturbations with the wavelength $O(\lambda _{0})$.

If one takes $\phi _{i}\gg O(1/\sqrt{M})$, a typical amplitude of the field $%
\sigma $ becomes much greater than $O(1)$ in Planck units, which means that
this field becomes capable of supporting chaotic inflation.\footnote{A closely related thought somewhat different possibility was mentioned by David Wands in his talk at the workshop after the conference \textquotedblleft The Next Chapter in
Einstein's Legacy\textquotedblright\ at the Yukawa Institute, Kyoto.} One may call this effect `the curvaton-inflaton transmutation.' 

In this regime
the field $\sigma $ typically serves as an inflaton rather than as a
curvaton.  Indeed, in this case we consider two fields, $\phi $ and $\sigma $, in a universe with a Hubble constant determined by a total energy density
of both fields. Their combined dynamics can be easily understood in the
approximation $H\approx const$. In this case $\phi =\phi _{i}\exp {\left( -{%
\frac{M^{2}t}{3H}}\right) }$, whereas $\sigma =\sigma _{0}\exp {\left( -{%
\frac{m^{2}t}{3H}}\right) }$. Obviously, for $m\ll M\lesssim H$ the field $%
\sigma $ practically does not change until inflationary stage driven by the
field $\phi $ is over. After that, the field $\sigma $ drives subsequent stages of 
inflation, producing usual inflaton perturbations with amplitude $O(m).$

Thus, if the field $\phi $ was originally greater than $O(1/\sqrt{M})$, then
the last stage of inflation is driven by the lightest field $\sigma $. This
happens even if the initial value of $\sigma $ is smaller than the Planck
value; see \cite{Kofman:1986wm} for a related discussion.

Until now we were assuming that the scalar field $\phi $ obeys classical
evolution described by Eq. (\ref{a5}). However, if $\phi _{i}\gg O(1/\sqrt{M}%
)$ then inflation is eternal \cite{Eternal,book}. Thus in eternal inflation
scenario based on the model with two fields, the last stage of inflation is
typically driven by the lightest field. In other words, the lightest field $%
\sigma$ typically is not a curvaton but an inflaton.

Moreover, even if eternal inflation is initially supported only by the field 
$\phi $, in some parts of the universe the quantum fluctuations bring this
field higher and higher, up to the Planck density. In those regions, the
Hubble constant grows up to the Planck value $H\sim 1$, and $\langle \sigma
^{2}\rangle $ eventually becomes 
\begin{equation}
\langle \sigma ^{2}\rangle =O(m^{-2})\ .  \label{1aes}
\end{equation}%
The energy density of the fluctuations of the field $\sigma $ also
approaches Planck density: ${\frac{1}{2}}m^{2}\langle \sigma ^{2}\rangle
=O(1)$. This is not surprising; in fact this always happens in the eternal
inflation scenario based on the simplest chaotic inflation which allows the
nearly Planckian energy density \cite{Eternal,LLM}. But this means that as a
result of quantum fluctuations produced during eternal inflation, the scalar
field $\sigma $ typically becomes much greater than the field $\phi $.
Indeed, the field $\sigma $ can take any value in the interval $|\sigma
|\lesssim 1/m$, which is much greater than $|\phi |\lesssim 1/M$, in Planck
units. Later on, the field $\phi $ rolls down faster, and the last stage of
inflation is eventually driven by the lighter field $\sigma $.

How these results depend on the choice of the inflaton potential $V\left(
\phi \right) $? Assuming that during inflation driven by the field $\phi $
the curvaton mass $m$ is negligibly small compared to the Hubble constant,
and the ratio $\frac{V^{2}\left( \phi _{i}\right) }{V^{\prime}}$ grows at
large $\phi $, we obtain from (\ref{a6}) the following estimate for $\langle
\sigma ^{2}\rangle $ at the end of the inflation:%
\begin{equation}
\langle \sigma ^{2}\rangle \sim \frac{V^{2}\left( \phi _{i}\right) \phi _{i}%
}{V^{\prime}\left( \phi _{i}\right) }.  \label{a9}
\end{equation}%
Therefore if at the beginning of inflation (i.e. at $\phi =\phi _{i}$) the
condition 
\begin{equation}
V^{2}\left( \phi \right) \phi >V^{\prime}\left( \phi \right)   \label{a10}
\end{equation}%
is satisfied, then $\langle \sigma ^{2}\rangle >1$, and the curvaton $\sigma 
$ becomes an inflaton. It is easy to verify that, for example, for the
power-law potential $V\sim \lambda \phi ^{n},$ the condition of
self-reproduction,%
\begin{equation}
V^{3/2}>V^{\prime},  \label{a11}
\end{equation}%
coincides with (\ref{a10}). Therefore all results which we derived for the
massive inflaton potential are valid in this case as well.

Thus we are coming to the conclusion that if eternal inflation takes place
in our model (or if the field $\sigma $ was greater than $O(1)$ from the
very beginning), then the scalar field $\sigma $ typically becomes extremely large and plays the role of
the inflaton rather than the curvaton. We will return to this conclusion and
its consequences in Section \ref{landscape}.

\subsection{ Curvaton with $m^{2}=\protect\alpha H^{2},$ $\protect\alpha < 1$%
}\label{alpha}

Another interesting case is when the curvaton mass squared during inflation
is proportional to the curvature $R=12H^{2}$ \cite{LM}. Mass terms $\sim
\alpha H^{2}$ appear in many models of supergravity, and also in the models
of the field $\sigma $ non-minimally coupled to gravity. Substituting $%
m^{2}=\alpha H^{2}=\frac{\alpha }{3}V$ in (\ref{a6}), we obtain%
\begin{equation}
\langle \sigma ^{2}\left( \phi \right) \rangle =\frac{1}{12\pi ^{2}}%
\int\limits_{\phi }^{\phi _{i}}\frac{V^{2}\left( \tilde{\phi}\right) }{V^{\prime}}\exp \left( -\frac{2\alpha }{3}\int\limits_{\phi }^{\tilde{\phi%
}}\frac{V}{V^{\prime}}d\bar{\phi}\right)\, d\tilde{\phi}  \label{a12}
\end{equation}%
For the case of the power-law potential $V$ the integral in (\ref{a12}) can
be calculated exactly. In particular, for $V=\frac{1}{2}M^{2}\phi ^{2},$ we
obtain%
\begin{equation}
\langle \sigma ^{2}\left( \phi \right) \rangle =\frac{M^{2}}{16\pi
^{2}\alpha }\left( \phi ^{2}+\frac{6}{\alpha }-\left( \phi _{i}^{2}+\frac{6}{%
\alpha }\right) e^{-\frac{\alpha }{6}\left( \phi _{i}^{2}-\phi ^{2}\right)
}\right)  \label{a13}
\end{equation}%
In the case $\alpha \ll \phi _{i}^{-2}$, this expression simplifies to 
\begin{equation}
\langle \sigma ^{2}\left( \phi \right) \rangle \simeq \frac{M^{2}}{192\pi
^{2}}\left( \phi _{i}^{4}-\phi ^{4}\right) \ .  \label{a14}
\end{equation}%
By taking $\phi \ll \phi_{i}$, we reproduce our previous result (\ref{a8})
for the amplitude of perturbations of the field $\sigma$ at the end of
inflation driven by the field $\phi$. Therefore, for $\alpha \ll M$, the
mass term $\alpha H^{2}$ is irrelevant at the self-reproduction scale $\phi
_{i}\sim M^{-1/2}$ and below. As a result, in this regime the curvaton $%
\sigma$ typically becomes an inflaton and drives the last stages of
inflation.

On the other hand, for sufficiently large $\alpha $, the behavior of the
curvaton filed in chaotic inflation is well described by the model with
constant $m$ and $H$ considered in Section \ref{pert}. Indeed, if $\alpha
\gg \phi _{i}^{-2}$, then, as soon the field $\phi $ becomes much smaller
than $\phi_{i} $, one can neglect the last term in (\ref{a13}), and the
result becomes%
\begin{equation}
\langle \sigma ^{2}\left( \phi \right) \rangle \simeq \frac{M^{2}}{16\pi
^{2}\alpha }\left( \phi ^{2}+\frac{6}{\alpha }\right) .  \label{a15}
\end{equation}%
For $\alpha \gg \phi ^{-2}$ we have%
\begin{equation}
\langle \sigma ^{2}\left( \phi \right) \rangle \simeq \frac{M^{2}\phi ^{2}}{%
16\pi ^{2}\alpha }\simeq \frac{3H^{4}\left( \phi \right) }{8\pi
^{2}m^{2}\left( \phi \right) } \ .  \label{a16}
\end{equation}%
Thus, the distribution of the field $\sigma $ behaves as a Bunch-Davies
distribution for a field with an adiabatically decreasing mass $m^{2}\approx
\alpha H^{2}$. A typical dispersion of the field $\sigma $ slowly decreases,
remaining of the order ${\frac{H^{2}(\phi )}{m}}\sim \alpha ^{{-1/2}}H\sim
\alpha ^{{-1/2}}M\phi $.

This result is easy to understand. As it was shown in Section \ref{pert},
the Bunch-Davies distribution (\ref{1ae}) is established during the time $%
t\sim {\frac{3H}{m^{2}}}$. All results obtained in Section \ref{pert} will
remain valid at all stages of inflation where the Hubble constant did not
change much during the time $t\sim {\frac{3H}{m^{2}}}={\frac{3}{\alpha H}}$.
One can show that this is equivalent to the condition 
\begin{equation}
\alpha \gg \left( {\frac{V^{\prime}}{V}}\right) ^{2}\sim \phi ^{-2} \ .
\end{equation}%
This adiabatic regime ends when $\phi^{2}$ becomes smaller than $\alpha^{-1}$%
. Thus, at the last stages of inflation Eq. (\ref{a14}) yields  
\begin{equation}
\langle \sigma ^{2}\rangle \simeq \frac{3M^{2}}{8\pi ^{2}\alpha ^{2}} \ .
\end{equation}%
Note that if $\alpha \gg M,$ then $\langle \sigma ^{2}\rangle <1$, and the
curvaton does not become an inflaton.

As we have shown in \cite{LM}, the slope of the spectrum of curvaton
perturbations produced in this model depends on the parameter $\alpha$. If $%
\alpha \ll 1$, the spectrum will be nearly flat, and perturbations will be
locally gaussian. If $\alpha$ is sufficiently large, then the typical size
of the cells of the curvaton web become small, the perturbations become
strongly nongaussian, and they may have blue spectrum.

We should emphasize again that the curvaton perturbations appear \textit{in
addition} to the usual adiabatic perturbations produced by the inflaton
fluctuations. By combining these two contributions, one can obtain adiabatic
perturbations with a controllable degree of nongaussianity. Also, since each
of these two contributions has its own spectral index, a combination of
these two spectra may look like a spectrum with a running spectral index.

\section{Curvaton web and stringy landscape}
\label{landscape}

In this section we will discuss an additional reason to be interested in the
curvaton models and the curvaton web.

With the development of inflationary scenario it became clear that the
universe may consist of exponentially large number of exponentially large
locally homogeneous domains of different types, which allowed us for the
first time to obtain a scientific justification of anthropic principle \cite%
{Linde:1984je,Eternal,LLM,mediocr,Garcia-Bellido:1994ci}. This possibility
became especially interesting when it was realized that string theory may
have an enormously large number of different vacua \cite%
{KKLT,BP,Susskind:2003kw,Douglas}.

One may wonder where is our place in the vast stringy landscape? The first
thing to say is that we simply cannot live in most of these vacua. For
example, we cannot live in a non-compactified 10D space even if its total
volume is extremely large. This is a simple superselection rule. One may try
to go further and evaluate the probability of emergence of life in a world
with different masses and coupling constants. This is what was traditionally
done by those who explored various consequences of anthropic principle prior
to invention of inflationary cosmology.

Inflation allows us to look at this problem from a different perspective.
One may assume that we are typical observers, and the total number
of observers is proportional to the total volume of the universe. Therefore
the probability to live in a part of the universe with given properties
depends not only on these properties, but also on the volume of this part of
the universe as compared to the volume of other parts. Since volume is
exponentially sensitive to parameters of inflationary theory and initial
conditions for inflation, investigation of the volume-weighted probability
distribution of different parts of the universe can give us a very powerful
tool for understanding of our place in the universe.

The volume-weighted probability distribution in inflationary cosmology was
introduced in \cite{Eternal,Goncharov:1987ir}. A detailed investigation of
this probability distribution and its applications to anthropic principle
was performed in \cite{LLM}, which was followed by many other works on this
subject, see e.g. \cite{mediocr,Garcia-Bellido:1994ci,Garriga:2005av,Tegmark:2004qd} and
references therein.

There are many problems related to the use of this probability distribution,
including the choice of a proper probability measure and a general issue of
its interpretation. These issues could seem abstract and
academical, but recently the situation changed. After the invention of the
mechanism of moduli stabilization in string theory and the development of
the concept of string theory landscape \cite{KKLT,BP,Susskind:2003kw,Douglas}
the problem of the choice of a most probable vacuum state became one of the
most important problems of modern physics. Many string theorists and
cosmologists began actively debating this question. It is not our goal to
discuss all of these issues in this paper. Here we will only
mention one aspect of this problem related to the curvaton web.

To explain the basic issue, let us consider, following \cite{mediocr,Feldstein:2005bm,Garriga:2005ee}, the simplest class of
inflationary models with a potential $V(\phi )={\frac{M^{2}}{2}}\phi ^{2}$,
which we discussed in Sect. \ref{chaotic}. During the rolling of this field
from its initial value $\phi _{i}$ the universe expands as $e^{\phi
_{i}^{2}/4}$, see Eq. (\ref{8aab}). Suppose inflation began at the Planck
boundary $V(\phi _{i})={\frac{M^{2}}{2}}\phi _{i}^{2}=1$. Then, ignoring the
issues of eternal inflation, the volume of the universe during inflation in
this scenario grows $\exp \left( {\frac{3}{2M^{2}}}\right) $ times. If
instead of that we assume that inflation begins at the eternal inflation
boundary $\phi _{i}\sim 1/\sqrt{M}$, one finds that the volume of the
universe grows by a factor of $\exp \left( O(M^{-1})\right) $. In either
case, the total volume of the universe after inflation grows exponentially
when $M$ decreases.

Thus if the inflaton mass $M$ can take different values in different parts of the universe, one may argue that the parts with a smaller $M$ will be exponentially bigger, so they will contain exponentially greater number of observers, and therefore a typical observer will live in a part of the universe with the smallest possible inflaton mass.

On the other hand, the amplitude of density perturbations produced in this
scenario is \cite{Pert,mukhbook,book} 
\begin{equation}
{\frac{\delta\rho}{\rho}} \sim {\frac{V^{3/2}}{V^{\prime}}} \sim {M\phi^{2}}
\sim M
\end{equation}
at the end of chaotic inflation, where $\phi = O(1)$. Since the volume of
the parts of the universe produced by inflation is exponentially sensitive
to the value of $M$, one may argue that it would be exponentially more
probable to live in the universe with the amplitude of density perturbations
much smaller than its present value.

This is a serious problem, which was especially clearly described in Ref. 
\cite{Garriga:2005ee} on the basis of a combined analysis of the probability distribution for the cosmological constant and the amplitude of density perturbations.

One should note that this problem appears mostly because of the
assumption that the probability to live in a given part of the universe is
proportional to the degree of inflation. While this is a reasonable assumption \cite{LLM}, its validity is not absolutely clear; see e.g. a discussion in 
\cite{Garcia-Bellido:1994ci,Freivogel:2005vv}. In particular, if one uses the
probability measure associated with the volume at the hypersurface of time
measured in units of $H^{-1}$, which was one of the two possible measures  advocated in \cite{LLM}, then the resulting probability distribution 
does not contain any factors of the type of $e^{\phi _{i}^{2}/4}$. In addition, we do not actually know whether it makes any sense to compare different causally disconnected parts of the universe with different parameters \cite{Garcia-Bellido:1994ci}.

These issues deserve a separate investigation, which goes beyond the scope of the present paper. Still, with all of these uncertainties in mind, the problem described above should not be discarded. In order to resolve this problem, it was suggested in \cite{mediocr} that
the cosmological perturbations of metric should be produced not by the inflaton field but by cosmic strings, in which case the amplitude of perturbations does not depend on $M$ and on the duration of inflation. This possibility by now is ruled out by cosmological observations. Therefore in \cite{Garriga:2005ee} it was proposed to use the curvaton mechanism of generation of density perturbations. The main idea is
that in the simplest versions of this scenario ${\frac{\delta\rho}{\rho_{\sigma}}}  \sim {\frac{H}{\pi\sigma}}$, see Eq. (\ref{locgauss}). This amplitude does not depend on the mass of
the inflaton field, so one can maximize the amount of inflation by
minimizing $M$, without affecting ${\frac{\delta\rho}{\rho}}$.

One of the problems with this proposal is that in this scenario the
probability to live in the universe with the amplitude of density
perturbations greater than some given value ${\frac{\delta\rho}{\rho}}$, and with the anthropic constraints imposed on the combination of  ${\frac{\delta\rho}{\rho}}$ and the cosmological constant $\Lambda$, is
proportional to $\left({\frac{\delta\rho}{\rho}}\right)^{2}$ \cite%
{Garriga:2005ee}. This result suggests that we should live in the universe with $%
{\frac{\delta\rho}{\rho}} \gg 10^{{-5}}$, which also disagrees with the observational data. In order to avoid this problem it
was suggested to consider more complicated, multi-component curvaton models \cite{Garriga:2005ee}.

In this respect, our investigation of the curvaton web brings both good and bad
news. The good news is that if one takes into account production of the
curvaton particles during reheating, then the amplitude of the curvaton
perturbations does not blow up at small $\sigma$. Instead  it has a
maximal value given by ${\frac{\delta\rho}{\rho}}_{\mathrm{max}} \sim {%
\frac{ H }{2\pi C}}$, see Eq. (\ref{locgausstot11ab}). This would be
consistent with observations for $C \sim 10^{4}H$.

The bad news is that this idea requires a specific carefully tuned mechanism of reheating after inflation. Moreover, this idea does not work at all in the simplest curvaton models. For example, as pointed out in Section \ref{chaotic}, in the eternal
inflation scenario in the model with the curvaton potential $%
m^{2}\sigma^{2}/2$ and the inflaton potential $M^{2}\phi^{2}/2$, the field $%
\sigma$ typically plays the role of the inflaton instead of the curvaton.

One can avert this conclusion by considering the field $\sigma$ with the
mass $m^{2} = m_{0}^{2} +\alpha H^{2}$ with a sufficiently large $\alpha$,
as we did in \cite{LM}; see Section \ref{alpha}. But this makes the model more complicated and prompts to look for alternatives ideas.

The simplest way to obtain $m^{2}=m_{0}^{2}+\alpha H^{2}$ is to add the term 
$-{\frac{\alpha }{24}}\sigma ^{2}R$ to the Lagrangian. But if one does it
for the curvaton, one may do it for the inflaton as well. By adding the term 
$-{\frac{\xi }{2}}\phi ^{2}R$ to the Einstein Lagrangian ${\frac{1}{2}}R$,
one would make it impossible for the field $\phi $ to be greater than $1/\sqrt{\xi }$ because the effective gravitational constant $G = (8\pi(1-\xi\phi^{2}))^{-1}$ becomes singular at $\phi =1/\sqrt{\xi }$. In this model the maximal amount of inflation can
be estimated by $e^{\phi _{i}^{2}/4}\sim e^{1/4\xi }$. This quantity no
longer depends on $M$, so the whole problem discussed in \cite%
{mediocr,Feldstein:2005bm,Garriga:2005ee} disappears.

More generally, this problem arises only because the models considered in 
\cite{mediocr,Feldstein:2005bm,Garriga:2005ee} depend on a single parameter,
such as the inflaton mass $M$, or a coupling constant $\lambda$ in the
potential $\lambda\phi^{n}$. As we just mentioned, this problem disappears
if we add a second parameter, $\xi$, and keep it fixed. Of course, one may
argue that this parameter must be minimized in its turn, in order to maximize the degree of inflation. But with each such step the reliability of the basic arguments  becomes less certain.

In this respect, it is quite instructive to consider a different modification of the theory. Let us add a constant $V_{0}$ to the potential $M^{2}\phi^{2}/2$, and
assume that inflation ends when the field $\phi$ reaches some point $\phi_{0}
$ with $M^{2}\phi_{0}^{2}/2 \ll V_{0}$, as it happens in the hybrid inflation scenario \cite{Hybrid}. In this
case, as before, the total duration of inflation will be determined by the behavior of
the universe at extremely large $\phi$, where $M^{2}\phi^{2}/2 \sim 1  \gg V_{0}$. This duration will be almost the same as
in the simple model $M^{2}\phi^{2}/2$. Meanwhile the amplitude of density
perturbations during the last 60 e-folds will be determined by inflation in the regime $
M^{2}\phi_{0}^{2}/2 \ll V_{0}$, and it will be proportional to $M^{-2}$ \cite{Hybrid}: 
\begin{equation}
{\frac{\delta\rho}{\rho}}  \sim {\frac{V^{3/2}}{V^{\prime}}} \sim {\frac{V_{0}^{3/2}}{M^{2}\phi_{0}}} \ .
\end{equation}
In this case one would have another problem: Long duration of inflation, as
before, favors small $M$, but now small $M$ lead to unacceptably {\it large}
density perturbations.

So what is the problem? Is it an unacceptably \textit{small} density
perturbations, as in the model $M^{2}\phi^{2}/2$, or an unacceptably \textit{large} density perturbations, as in the model $M^{2}\phi^{2}/2 +V_{0}$?

The root of this ambiguity is easy to understand. The degree of
inflation in this scenario is determined by the shape of the potential at
extremely large $\phi$, whereas the amplitude of density perturbations is
determined by the shape of the potential at relatively small $\phi$. 
The problem discussed in \cite{mediocr,Feldstein:2005bm,Garriga:2005ee} appears
only if one believes that the shape of the inflaton potential is determined
by a single parameter and does not change during an incredibly large
interval of variation of the inflaton field. We do not think that this
requirement is generically satisfied by inflationary models in stringy
landscape. 

The simplest way to address this problem is to maximize the total amount of inflation by properly tuning the part of the potential describing the main part of the process of inflation (which corresponds to $\phi \gg O(10)$ in the simplest chaotic inflation models), and then tune the part describing the last 60 e-folds of inflation by tuning the remaining part of the potential ($\phi \lesssim O(10)$ in our example). As our example shows, one can easily tune the amplitude of perturbations of metric anywhere in the interval from $0$ to $O(1)$ without altering the early stage of inflation which is responsible for the total degree of inflation.

Therefore we believe that it would be somewhat premature to use anthropic considerations for justification of the curvaton scenario. 

\section{Conclusions}
In this paper we discussed the global structure of the universe in the simplest curvaton scenario proposed in our paper \cite{LM}. This scenario is based on the theory of two massive scalar fields, the inflaton $\phi$ and the curvaton $\sigma$, interacting with each other only gravitationally. We have shown that stochastic processes during eternal inflation in this scenario divide the universe into many exponentially large domains containing either positive or negative values of the curvaton field $\sigma$. The amplitude of the curvaton perturbations in this scenario depends on the process of reheating after inflation and has a rather nontrivial behavior as a function of the distance from the walls $\sigma = 0$ separating the domains with positive and negative $\sigma$. We called the resulting structure, which was shown in Figs. \ref{fig:Figa}, \ref{fig:Figb}, `the curvaton web.'

Depending on the parameters of the model and on our own position with respect to the curvaton web, we may live in a universe with strongly nongaussian adiabatic perturbations, or in the universe with locally gaussian adiabatic perturbations with the amplitude depending on our position, or in the universe dominated by the usual gaussian inflaton perturbations with a small admixture of nongaussian curvaton perturbations. In the last case, one may relax many constraints imposed by the requirement of a secondary reheating and creation of dark matter and baryons after the curvaton decay, and instead of the adiabatic perturbations produced by the decaying curvaton study a small admixture of isocurvature nongaussian perturbations. Moreover, in different parts of the universe the same scalar fields may play either the role of the curvaton or the role of the inflation; the effect which we called `the curvaton-inflaton transmutation.' 

This example shows that in addition to a {\it discrete} spectrum of possibilities, which appear after compactification in string landscape scenario, one may have a {\it continuous} spectrum of possibilities, which appear because of a complicated dynamics in the early universe. The availability of so many new possibilities is encouraging, but also rather  disturbing. The curvaton scenario is much more complicated than the usual scenario where the adiabatic perturbations are produced by the inflaton fluctuations, so we believe that it should be used sparingly. On the other hand, in certain situations the curvaton scenario can be quite useful. In particular, it provides by far the simplest way to obtain a controllable amount of nongaussian perturbations of metric  \cite{LM}. It has many other interesting and unusual features, some of which were discussed in this paper, so it certainly deserves further investigation.

\begin{acknowledgments}
This paper was initiated by discussions during the conference ``The Next
Chapter in Einstein's Legacy'' at the Yukawa Institute, Kyoto, and the
subsequent workshop. We are very are grateful to the organizers of this
conference for hospitality. We are especially grateful to Lev Kofman, Misao Sasaki and Alex Vilenkin for valuable discussions. We are especially grateful to David Lyth for many enlightening comments. The work by A.L. was supported by NSF grant
PHY-0244728 and by Kyoto University.
\end{acknowledgments}

\end{document}